\documentclass[aps,prd,nofootinbib,amsmath,amsfonts,preprintnumbers,notitlepage,10pt,english]{revtex4-1}
\usepackage{hyperref}
\usepackage{amsmath}
\usepackage{amssymb}
\usepackage{babel}
\usepackage{graphicx}

\newcommand{\f}[2]{\frac{#1}{#2}}
\newcommand{\mk}[1]{\left( #1 \right)}
\newcommand{\kk}[1]{\left[ #1 \right]}

\newcommand{\be}{\begin{equation}}
\newcommand{\ee}{\end{equation}}
\def\Mpl{M_{\rm Pl}}


\begin{document}

\title{
Consistency relation for $R^p$ inflation}

\author{Hayato Motohashi}
\affiliation{Kavli Institute for Cosmological Physics, The University of Chicago, 
Chicago, Illinois 60637, U.S.A.}

\begin{abstract}%%%%%%%%%%%%%%%%%%%%%%%%%%%%%%%%%%%%%%%%%
We consider $R^p$ inflation
with $p \approx 2$, allowing small deviation from $R^2$ inflation.
Using the inflaton potential in the Einstein frame, we construct a consistency relation between the scalar spectral index, the tensor-to-scalar ratio, as well as the running of the scalar spectral index, which will be useful to constrain a deviation from $R^2$ inflation in future observations.
\end{abstract}
% 04.50.Kd Modified theories of gravity
% 98.80.Cq Particle-theory and field-theory models of the early Universe (including cosmic pancakes, cosmic strings, chaotic phenomena, inflationary universe, etc.) 
% 98.80.-k Cosmology

\date{\today}

\maketitle

\section{Introduction}%%%%%%%%%%%%%%%%%%%%%%%%%%%%%%%%%%%%%%%%%

The first self-consistent model of inflation is $R^2$ inflation proposed by Starobinsky in 1980~\cite{Starobinsky:1980te},
where $R$ is the Ricci curvature. 
This model incorporates a graceful exit to the radiation-dominated stage via a period of reheating, where the standard model particles are created through the oscillatory decay of the inflaton, or dubbed the scalaron~\cite{Vilenkin:1985md,Mijic:1986iv,Ford:1986sy}.
The predictions of $R^2$ inflation for the spectra of primordial density perturbations and gravitational waves remain in agreement with the most recent high-precision data of the cosmic microwave background (CMB)~\cite{Hinshaw:2012aka,Ade:2013uln}.
In March 2014, BICEP2 announced the detection of B-mode polarization at degree angular scales in the CMB, and the amplitude of the tensor-to-scalar-ratio
is as large as $r=0.20^{+0.07}_{-0.05}$~\cite{Ade:2014xna}, which is 
in tension with previous data as well as the prediction of $R^2$ inflation.
However, it is still unclear if the signal is of primordial origin, due to an unknown amplitude of foreground dust emission~\cite{Adam:2014bub}.
In light of this, $R^2$ inflation is still consistent with the recent data and upcoming data may allow us to pin down the inflationary model of our universe.

In addition to inflation, the $R^2$ term play a different role in the context of $f(R)$ gravity for the late-time acceleration.
By choosing a suitable functional form of $f(R)$, $f(R)$ gravity can mimic the expansion history of the concordance $\Lambda$CDM model without a cosmological constant~\cite{Hu:2007nk,Appleby:2007vb,Starobinsky:2007hu}. Observationally, a key to distinguish $f(R)$ gravity from the $\Lambda$CDM model is the expansion history and the growth of the large-scale structure, which are conveniently parametrized by the equation-of-state parameter $w$ for dark energy and the growth index $\gamma$, respectively, because both parameters remain constant in the $\Lambda$CDM model, namely, $w=-1$ and $\gamma=0.55$, while they are dynamical in $f(R)$ gravity~\cite{Hu:2007nk,Motohashi:2010tb,Motohashi:2011wy,Gannouji:2008wt,Motohashi:2009qn,Tsujikawa:2009ku,Motohashi:2010sj}. 
In particular, it is interesting that $f(R)$ gravity allows a 1 eV sterile neutrino~\cite{Motohashi:2012wc}, whose existence has been suggested by recent neutrino oscillation experiments but is in tension with vanilla $\Lambda$CDM.
However, the $f(R)$ models for the late-time acceleration suffer from singularity problems, where the scalaron mass and Ricci curvature diverge quickly in the past~\cite{Starobinsky:2007hu,Tsujikawa:2007xu,Appleby:2008tv,Frolov:2008uf,Kobayashi:2008tq}.
These problems are solved if we add $R^2$ term
\cite{Appleby:2009uf}. The resultant combined $f(R)$ model incorporates inflation and the late-time acceleration. In the combined model, inflationary dynamics is still the same as $R^2$ inflation, while differences show up in reheating phase dominated by the kinetic energy of the scalaron~\cite{Motohashi:2012tt}, which enhances the tensor power spectrum~\cite{Nishizawa:2014zra}.

The $R^2$ model is thus attractive in the sense that   
it is currently one of the leading candidates for inflation and 
it cures singularity problems when combined with $f(R)$ models for the late-time acceleration. 
Although it is simple and powerful, with progress in observational accuracy, we can test for further complexity. Similar to the generalization from a scale-invariant spectrum to a nonzero tilt,
we may be forced to consider a small deviation from $R^2$ inflation.
Specifically, tiny tensor-to-scalar ratio for $R^2$ inflation, namely, $r\simeq 0.003$ for 60 e-folds, motivates us to consider a deviation from $R^2$ inflation.
We are poised to possibly obtain strong constraints on $r$ from joint analysis of Planck and BICEP2 data, along with future experiments.
It is therefore interesting to consider the possibility to generate larger value of $r$ based on the $R^2$ model.

In order to establish a way to measure a deviation from $R^2$ inflation, we investigate $R^p$ inflation in the present paper, where $p\approx 2$ and is not an integer, allowing small deviation from $p=2$. 
The $R^p$ Lagrangian was originally considered in the context of higher derivative theories~\cite{Schmidt:1989zz,Maeda:1988ab} and then applied to inflation~\cite{Muller:1989rp,Gottlober:1992rg} (see also \cite{DeFelice:2010aj,Martin:2014vha,Martin:2013nzq} for recent review), which provides a simple and economical generalization of $R^2$ inflation. Recently, $R^p$ inflation has been focused in the context of the generation of large $r$. 
It has been emphasized that for $p$ slightly smaller than $2$, the tensor-to-scalar ratio can be enhanced relative to the original $R^2$ model~\cite{Codello:2014sua} (see also \cite{Martin:2014lra}).
A combined $f(R)$ model based on the $R^p$ model has also been proposed~\cite{Artymowski:2014gea}.
Not only is it of phenomenological interest, the $R^p$ action is also theoretically motivated because one-loop corrections to the $R^2$ action could give a correction to the power of the Ricci scalar~\cite{Codello:2014sua,Ben-Dayan:2014isa,Rinaldi:2014gua}. 
A deformation of the $R^2$ action that mimics higher-loop corrections is considered in~\cite{Rinaldi:2014gha}. A relevance to Higgs inflation is considered in~\cite{Costa:2014lta,Chakravarty:2014yda}.

However, its prediction to the scalar spectral index $n_s$, the tensor-to-scalar ratio $r$, as well as the running of the scalar spectral index $\alpha\equiv dn_s/d\ln k$ is not well formulated. In particular, some of the previous results provide different results for the prediction of $n_s$ and $r$. 
Further, the running of the scalar spectral index in the model has not been discussed in the literature.
The aim of the present paper is to resolve these issues and present a consistency relation for $R^p$ inflation by using the inflaton potential in the Einstein frame. 
We consider not only  
the scalar spectral index and the tensor-to-scalar ratio, but also the running of the scalar spectral index. 
We derive a handy expression for these inflationary observables, which will be useful to constrain a deviation from $R^2$ inflation in future observations.

The organization of the paper is as follows.
In Sec.~\ref{sec-inf}, we explore the background dynamics of the inflationary expansion in $R^p$ inflation. We write down the inflaton potential for general $p$ in the Einstein frame and the slow-roll parameters in terms of the derivatives of the potential.
In Sec.~\ref{sec-cr}, we derive a consistency relation between the inflationary observables, with which we can constrain the model.
We conclude in Sec.~\ref{sec-cn}.
Throughout the paper, we will work in natural units where $c=1$, and the metric signature is $(-+++)$.

\section{$R^p$ inflation}%%%%%%%%%%%%%%%%%%%%%%%%%%%%%%%%%%%%%%%%%
\label{sec-inf}

Let us start with a general $f(R)$ and write down equations of motion in the Einstein frame.
We consider 
\be S=\int d^4x \sqrt{-g} \f{\Mpl^2}{2}f(R),
\ee
where $\Mpl\equiv (8\pi G)^{-1/2}$ is the reduced Planck mass.
By using the conformal transformation $g^E_{\mu\nu}\equiv F(R) g_{\mu\nu}$ with defining the scalaron field $\phi$ by $F(R) \equiv f'(R)\equiv e^{\sqrt{\f{2}{3}}\f{\phi}{\Mpl}}$, we can recast the action as
\be S=\int d^4x \sqrt{-g_E} \kk{\f{M_{\rm{Pl}}^2}{2} R_E-\f{1}{2}g_E^{\mu\nu} \partial_\mu \phi \partial_\nu \phi-V(\phi)}, \ee
where the potential is given by
\be \label{pote} V(\phi)=\f{\Mpl^2}{2}\f{\chi F(\chi)-f(\chi)}{F(\chi){}^2}. \ee
Here, $\chi=\chi(\phi)$ is a solution for $F(\chi) = e^{\sqrt{\f{2}{3}}\f{\phi}{\Mpl}}$, and thus $f(\chi)$ and $F(\chi)$ are determined for each $\phi$.
The time and the scale factor in the Jordan frame and Einstein frame relate through
\begin{align}
\label{conft} dt_E =\sqrt{F}dt,\quad
a_E =\sqrt{F}a,
\end{align}
and thus the Hubble parameter in the Einstein frame is given by
\be \label{hub} H_E=\f{H}{\sqrt{F}}\mk{1+\f{\dot F}{2HF}}, \ee
where a dot implies the derivative with respect to the time $t$ in the Jordan frame.
The Einstein equation reads
\begin{align}
\label{fe1} 3\Mpl^2 H_E^2 &= \f{1}{2}\mk{\f{d\phi}{dt_E}}^2+V, \\
\label{fe2} -2\Mpl^2 \f{dH_E}{dt_E} &= \mk{\f{d\phi}{dt_E}}^2, 
\end{align}
and the equation of motion for the scalaron is given by
\be
\label{fe3} \f{d^2\phi}{dt_E^2}+3H_E\f{d\phi}{dt_E}+V_\phi=0,
\ee
where $V_\phi\equiv \partial V/\partial \phi$.

For the rest of the paper, we focus on the following model:
\be \label{Rpmodel} f(R)=R+\lambda R^p. \ee
The parameter $p$ is not necessarily an integer in general, and $\lambda$ has mass dimension $(2-p)$. 
In this model, the potential \eqref{pote} can be explicitly written in terms of $\phi$ as
\be \label{pot} V(\phi)=V_0 e^{-2\sqrt{\f{2}{3}}\f{\phi}{\Mpl} } \mk{e^{\sqrt{\f{2}{3}}\f{\phi}{\Mpl}} -1}^{\f{p}{p-1}} \ee
with $V_0\equiv \f{\Mpl^2}{2}(p-1)p^{p/(1-p)} \lambda^{1/(1-p)} $.
Note that for $p=2$ and $\lambda=1/(6M^2)$, the potential \eqref{pot} recovers the potential for $R^2$ inflation: 
\be \label{pot2} V(\phi)=\f{3}{4}M^2\Mpl^2 (1- e^{-\sqrt{\f{2}{3}}\f{\phi}{\Mpl}} )^2, \ee
where the energy scale is normalized as $M\simeq 10^{13}$ GeV from the amplitude of observed power spectrum for the primordial perturbations.

\begin{figure}[t]
	\centering
	\includegraphics[width=100mm]{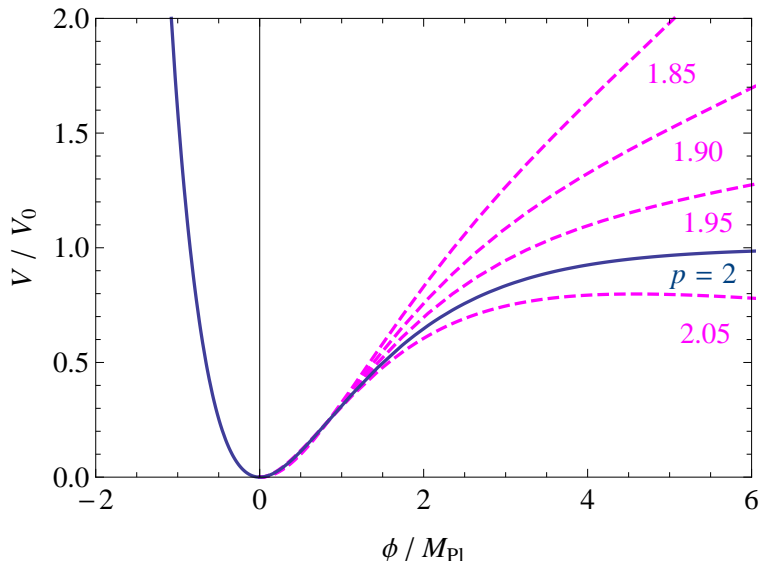}
	\caption{
		Potential for $R^p$ inflation with $p=2$ (blue solid), and $1.85$, $1.90$, $1.95$, $2.05$ (magenta dashed). }
	\label{fig:pot}
\end{figure}

In Fig.~\ref{fig:pot}, we present the potential \eqref{pot} for various $p$ around $p=2$.
The scalaron rolls slowly on the potential at $\phi>0$, and leads the inflationary expansion.
While the potential for $p=2$ asymptotically approaches to a constant value $V_0$ for large $\phi$, the potential for $p\lesssim 2$ continuously grows.
Therefore, the potential for $p\lesssim 2$ is steeper than $p=2$, and this leads to larger tensor-to-scalar ratio relative to $R^2$ inflation, as we shall see later.
For $p>2$, the potential \eqref{pot} has a maximum at $\phi=\Mpl\sqrt{\f{3}{2}}\ln \kk{\f{2(p-1)}{p-2}}\equiv \phi_m$ and approaches to $0$ for large $\phi$. For instance, $\phi_m/\Mpl\simeq 4.58$ for $p=2.05$. Therefore, inflation can take place at either of $0<\phi<\phi_m$ or $\phi>\phi_m$. We are interested in the former case to see a deviation from $R^2$ inflation, and do not consider the latter case, which leads to a completely different scenario from $R^2$ inflation.

%==================== Figure ====================
%==================== Figure ====================

We define the slow-roll parameters for the potential in the Einstein frame as
\begin{align}
\label{slow} \epsilon \equiv \f{\Mpl^2}{2}\mk{\f{V_\phi}{V}}^2, \quad 
\eta \equiv \Mpl^2\f{V_{\phi\phi}}{V}, \quad
\xi \equiv \Mpl^4\f{V_\phi V_{\phi\phi\phi}}{V^2} .
\end{align}
Under the slow roll approximation, \eqref{fe1} -- \eqref{fe3} read
\begin{align}
\label{slowapp} H_E\simeq \f{\sqrt{V}}{\sqrt{3}\Mpl}, \quad
\f{dH_E}{dt_E}\simeq -\f{V_\phi^2}{6V}, \quad
\f{d\phi}{dt_E}\simeq -\f{\Mpl V_\phi}{\sqrt{3V}}. 
\end{align}

During slow-roll regime, the scale factor in the Einstein frame undergoes a quasi-de Sitter expansion. 
From $|\dot F/(HF)| \simeq 2\sqrt{\epsilon/3} \ll 1$, $F$ remains approximately constant during the slow-roll regime. 
Hence, from \eqref{conft} the scale factor and time in the Einstein frame are identical to those in the Jordan frame up to a constant factor.
Consequently, the quasi-de Sitter expansion takes place in both frame.
The number of e-folds 
between an initial time $t_{Ei}$ and $t_{E}$ is given by
\be \label{efolds} N_E\equiv \int^{t_{E}}_{t_{Ei}} H_E dt_E 
\simeq \f{1}{\Mpl^2} \int^{\phi_i}_{\phi} \f{V}{V_\phi}d\phi.
\ee
Note that from \eqref{conft} and \eqref{hub} $H_E dt_E=Hdt[1+\dot F/(2HF)] \simeq Hdt$ during the slow-roll regime and therefore $N_E \simeq N$.
Armed with these equivalences between quantities in the Jordan frame and Einstein frame during inflation, 
we omit the subscript $E$ for the following and continue to explore the inflationary dynamics in the Einstein frame.

Before proceeding to detailed analysis for $p=2$ and general $p$, let us here clarify the differences of the potential in the previous works. 
In~\cite{Costa:2014lta}, the authors consider $R^p$ model \eqref{Rpmodel} at first but eventually investigate the potential $V\propto (1-\gamma e^{-\beta \phi})$ with $\beta$ and $\gamma$ as free parameters. This potential is obviously different from the potential \eqref{pot} in $R^p$ inflation because their potential approaches constant for large $\phi$. They show that $\epsilon \ll |\eta|$ always holds, and $n_s$ depends only on e-folds while $r$ depends on the model parameters and e-folds. As we shall see below, these points are incompatible with $R^p$ inflation.

In~\cite{Chakravarty:2014yda}, the authors also start from $R^p$ model \eqref{Rpmodel} but arrive the potential $V\propto  e^{\f{2-p}{p-1} \sqrt{\f{2}{3}} \f{\phi}{\Mpl}}$, assuming $\phi/\Mpl \gg \sqrt{3/2} \simeq 1.22$. However, as we shall see, a field value which we are interested in is the same order of $1.22$. In particular, their approximation breaks down as $p\to 2$, because the field value of our interest becomes closer to $1.22$. Actually, their $n_s$ and $r$ does not recover $R^2$ inflation. Therefore, we cannot use their result if we want to consider small deviation from $R^2$ inflation.

Thus, although both works are motivated by $R^p$ inflation, they did not investigate $R^p$ inflation itself. Rather, they investigated the potential $V\propto (1-\gamma e^{-\beta \phi})$ and $V\propto e^{\f{2-p}{p-1} \sqrt{\f{2}{3}} \f{\phi}{\Mpl}}$, respectively, both of which cannot be used as an asymptotic form of the potential \eqref{pot} of $R^p$ inflation. On the other hand, our analysis is based on the potential \eqref{pot} without any approximation.

\subsection{$p=2$}%%%%%%%%%%%%%%%%%%%%%%%%%%%%%%%%%%%%%%%%%
\label{sec-inf2}

First, let us focus on the case with $p=2$. 
The slow-roll parameters \eqref{slow} for the potential \eqref{pot2} are given by
\begin{align}
\epsilon &= \f{4}{3(e^{\sqrt{\f{2}{3}}\f{\phi}{\Mpl} } -1)^2} , \notag \\
\eta &= -\f{4(e^{\sqrt{\f{2}{3}}\f{\phi}{\Mpl} } -2) }{3(e^{\sqrt{\f{2}{3}}\f{\phi}{\Mpl} } -1)^2} , \notag \\
\label{slow2} \xi &= \f{16(e^{\sqrt{\f{2}{3}}\f{\phi}{\Mpl} } -4) }{9(e^{\sqrt{\f{2}{3}}\f{\phi}{\Mpl} } -1)^3}.
\end{align}
Thus the slow-roll parameters relate each other through $\phi$, and we can derive the following relation between them:
\begin{align}
\eta&= -\f{2\sqrt{\epsilon}}{\sqrt{3}}+\epsilon, \notag\\
\label{conslow2} \xi&= \f{4}{3}\epsilon-2\sqrt{3}\epsilon^{3/2}.
\end{align}
Note that these relations are derived by only using the form of the potential. They hold exactly, regardless of the appearance of the slow-roll parameters.
As we shall see later, it is when we convert these relations into a consistency relation between inflationary observables 
that we need the slow-roll approximation.

For $\phi>\Mpl$, the slow roll parameters are suppressed as $\epsilon$, $\xi \sim e^{-2\sqrt{\f{2}{3}}\f{\phi}{\Mpl} }$, and $|\eta| \sim e^{-\sqrt{\f{2}{3}}\f{\phi}{\Mpl} }$.
It is worthwhile to note that the hierarchy between the slow-roll parameters is not $1\gg \epsilon \sim |\eta| \gg \xi$ like $\phi^2$ inflation, but $1\gg |\eta| \gg \epsilon \sim \xi$, which leads to a tiny tensor-to-scalar ratio.

\begin{figure}[t]
	\centering
	\includegraphics[width=85mm]{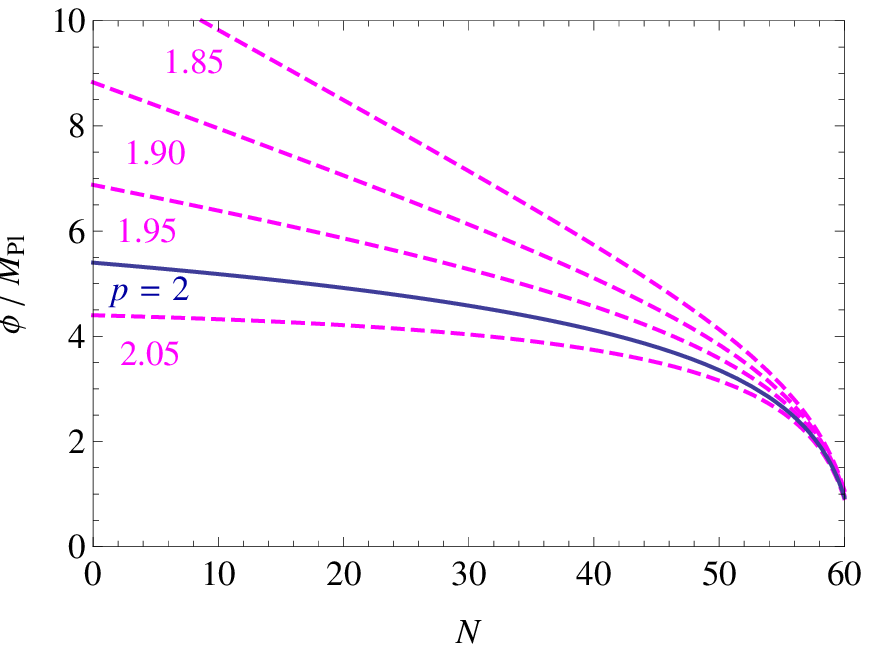}
	\includegraphics[width=85mm]{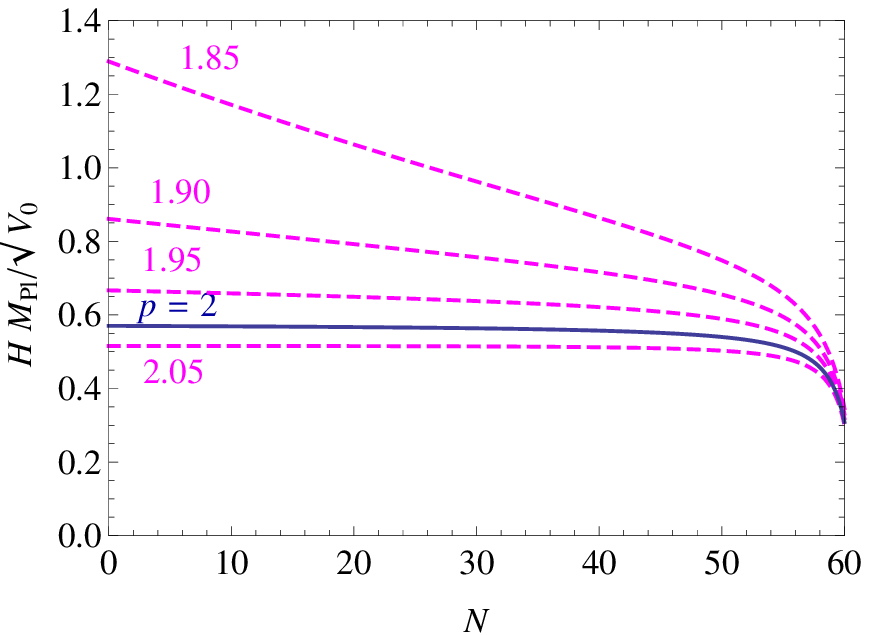}
	\caption{
		Time evolution of the scalaron $\phi$ and the Hubble parameter $H$ for $p=2$ (blue solid), and $1.85$, $1.90$, $1.95$, $2.05$ (magenta dashed).  
	}
	\label{fig:dyn}
\end{figure}

If we define the end of inflation by $\epsilon=1$, a field value at the end of inflation $\phi_f$ is given by
$\phi_f/\Mpl
%=\sqrt{3/2}\ln(1+2/\sqrt{3})
\simeq 0.940$. 
From \eqref{efolds}, we obtain the e-folds between $\phi_i$ and $\phi$ as 
\be N(\phi)=\f{3}{4}\mk{ e^{\sqrt{\f{2}{3}}\f{\phi_i}{\Mpl} } - e^{\sqrt{\f{2}{3}}\f{\phi}{\Mpl} } }, \ee
where we neglect a linear term of $(\phi-\phi_i)$, which gives a few percent correction.
We can solve this equation for $\phi$,
\be \f{\phi(N)}{\Mpl}=\sqrt{\f{3}{2}} \ln \mk{ e^{\sqrt{\f{2}{3}}\f{\phi_i}{\Mpl} } - \f{4}{3}N }, \ee
and using the slow-roll equation \eqref{slowapp} with the potential \eqref{pot2}, the Hubble parameter is given by
\be \f{H(N)}{\sqrt{V_0}/\Mpl}=\f{1}{\sqrt{3}} \kk{1- \mk{ e^{\sqrt{\f{2}{3}}\f{\phi_i}{\Mpl} } - \f{4}{3}N }^{-1} }, \ee
which are presented as a blue solid line in Fig.~\ref{fig:dyn}.

%==================== Figure ====================
%==================== Figure ====================

If we require the total e-folds $N_k\equiv N(\phi_f)=60$, we obtain $\phi_i/\Mpl
%=\sqrt{3/2}\ln(1+2/\sqrt{3}+(4/3)\times 60)
\simeq 5.40$.
Therefore, $N_k\simeq \f{3}{4}e^{\sqrt{\f{2}{3}}\f{\phi_i}{\Mpl} } $, and at the leading order of $N_k$, the slow roll parameters \eqref{slow2} at $\phi\simeq \phi_i$ are expressed as 
\begin{align}
\label{slow2n} \epsilon = \f{3}{4N_k^2},\quad
\eta = -\f{1}{N_k},\quad
\xi = \f{1}{N_k^2}.
\end{align}

\subsection{$p\approx 2$}%%%%%%%%%%%%%%%%%%%%%%%%%%%%%%%%%%%%%%%%%
\label{sec-infp}

We proceed to a general case with $p\approx 2$.
The slow-roll parameters \eqref{slow} 
are given by
\begin{align}
\epsilon 
&=\f{ [(2-p)F+2(p-1)]^2 }{ 3(p-1)^2(F-1)^2 }, \notag \\
\eta 
&=\f{ 2[ (2-p)^2F^2-(p-1)(5p-8)F+4(p-1)^2 ] }{ 3(p-1)^2(F-1)^2 },\notag \\
\label{slowpf} \xi 
&=\f{ 4[(2-p)F+2(p-1)] [ (2-p)^3F^3+(p-1)(2p-3)(5p-8)F^2-(p-1)^2(17p-24)F+8(p-1)^3 ] }{ 9(p-1)^4(F-1)^4 } , 
\end{align}
where $F\equiv e^{\sqrt{\f{2}{3}}\f{\phi}{\Mpl} }$ as we defined the above.
We can confirm that 
For $p=2$, \eqref{slowpf} reproduces \eqref{slow2}.
We can erase $F$ from these equations and obtain 
\begin{align}
&-4(2-p)+2(3p-4)\sqrt{3\epsilon}-6\epsilon+3p\eta=0,\notag\\
&-2(2-p)(3p-4)\sqrt{3\epsilon}+3(7p^2-24p+24)\epsilon-9\sqrt{3}(3p-4)\epsilon^{3/2}+9(2-p)\epsilon^2-\f{9}{4}p^2\xi=0.
\end{align}
Again, these relations hold without the slow-roll approximation.

The field value at the end of inflation $\epsilon=1$ is given by 
\be \f{\phi_f}{\Mpl} = \sqrt{\f{3}{2}} \ln \kk{\f{(2+\sqrt{3})(p-1)}{(1+\sqrt{3})p-(2+\sqrt{3})}}. \ee
For instance, $\phi_f/\Mpl
\simeq 0.907$, $0.978$, $1.02$, $1.07$ for $p=2.05$, $1.95$, $1.90$, $1.85$, respectively.

The number of e-folds between $\phi_i$ and $\phi$ given by \eqref{efolds} reads
\be N(\phi) = \f{3p}{4(2-p)}\ln \kk{\f{ (2-p) e^{\sqrt{\f{2}{3}}\f{\phi_i}{\Mpl} } +2(p-1) }{ (2-p) e^{\sqrt{\f{2}{3}}\f{\phi}{\Mpl} } +2(p-1) }}.
\ee
Then we obtain 
\be
\f{\phi(N)}{\Mpl} = \sqrt{\f{3}{2}} \ln \kk{ E^{-1}\mk{ e^{\sqrt{\f{2}{3}}\f{\phi_i}{\Mpl}}+\f{2(p-1)}{2-p} } - \f{2(p-1)}{2-p} }. \ee
From \eqref{slowapp}, the Hubble parameter is given by 
\be
\f{H(N)}{\sqrt{V_0}/\Mpl} = \f{1}{\sqrt{3}} \kk{ E^{-1}\mk{ e^{\sqrt{\f{2}{3}}\f{\phi_i}{\Mpl}}+\f{2(p-1)}{2-p} } - \f{p}{2-p} }^{\f{p}{2(p-1)}}
\kk{ E^{-1}\mk{ e^{\sqrt{\f{2}{3}}\f{\phi_i}{\Mpl}}+\f{2(p-1)}{2-p} } - \f{2(p-1)}{2-p} }^{-1}, \ee
where $E(N)\equiv e^{4(2-p)N/(3p)}$.
We present the time evolution of the scalaron and the Hubble parameter for $p=2.05$, $1.95$, $1.90$, $1.85$ by magenta dashed lines in Fig.~\ref{fig:dyn}.
As expected, the scalaron rolls down faster for $p \lesssim 2$.

By setting $N_k\equiv N(\phi_f)=60$, we obtain 
\be \f{\phi_i}{\Mpl} = \sqrt{\f{3}{2}} \ln \kk{ E_k \mk{ e^{\sqrt{\f{2}{3}}\f{\phi_f}{\Mpl}}+\f{2(p-1)}{2-p} } - \f{2(p-1)}{2-p} }, \ee
where $E_k\equiv e^{4(2-p)N_k/(3p)}$. For instance, $\phi_i/\Mpl\simeq 4.40$, $6.88$, $8.83$, $11.2$ for $p=2.05$, $1.95$, $1.90$, $1.85$, respectively.
Therefore, for $N_k$ we can neglect the contribution from $\phi_f$ and end up with
\be  N_k \simeq \f{3p}{4(2-p)}\ln \kk{ \f{(2-p)}{2(p-1)} e^{\sqrt{\f{2}{3}}\f{\phi_i}{\Mpl} } +1 }.
\ee
By taking the limit of $p\to 2$, we recover $N_k=\f{3}{4}e^{\sqrt{\f{2}{3}}\f{\phi_i}{\Mpl} }$.

By substituting $F=2(E_k-1)(p-1)/(2-p)$, we obtain the slow-roll parameters \eqref{slowpf} at $\phi\simeq \phi_i$ as
\begin{align}
\epsilon &= \f{4 E_k^2 (2-p)^2}{3 [2 (p-1)E_k - p]^2}, \notag \\
\eta &= \f{4 (2-p) [2 (2-p)E_k^2 - pE_k + p] }{3 [ 2 (p-1)E_k-p]^2}, \notag \\
\label{slowpn} \xi &= \f{16 E_k (2-p)^2 [4 (2-p)^2 E_k^3 + 2 p(4 p-7)E_k^2 -p(11p-18)E_k + p(3p-4)]}{9 [2 (p-1)E_k-p]^4}.
\end{align}
Taking the limit $p\to 2$, we can recover the results in $R^2$ inflation.

\begin{figure}[t]
\centering
\includegraphics[width=85mm]{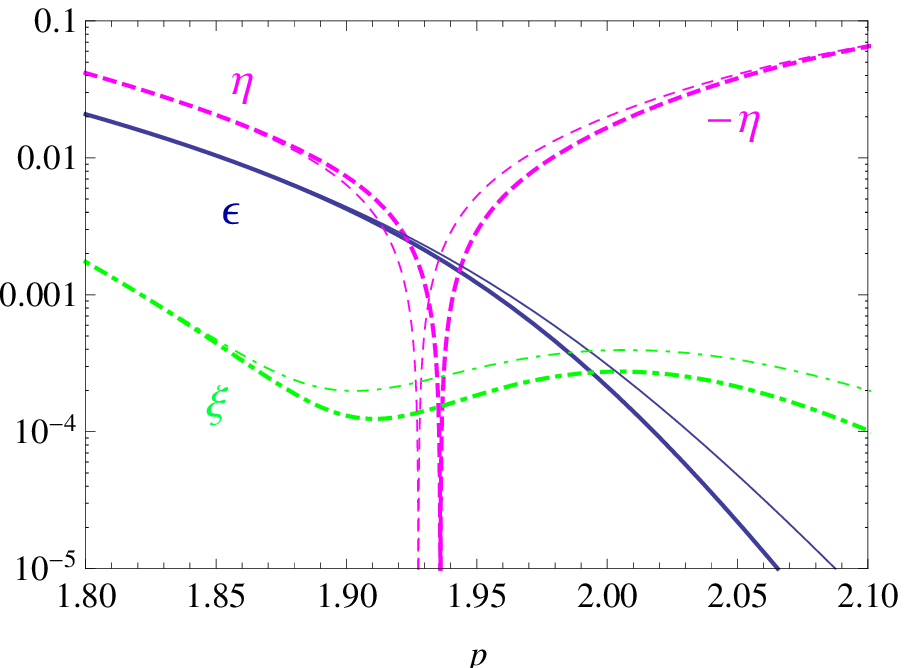}
\includegraphics[width=85mm]{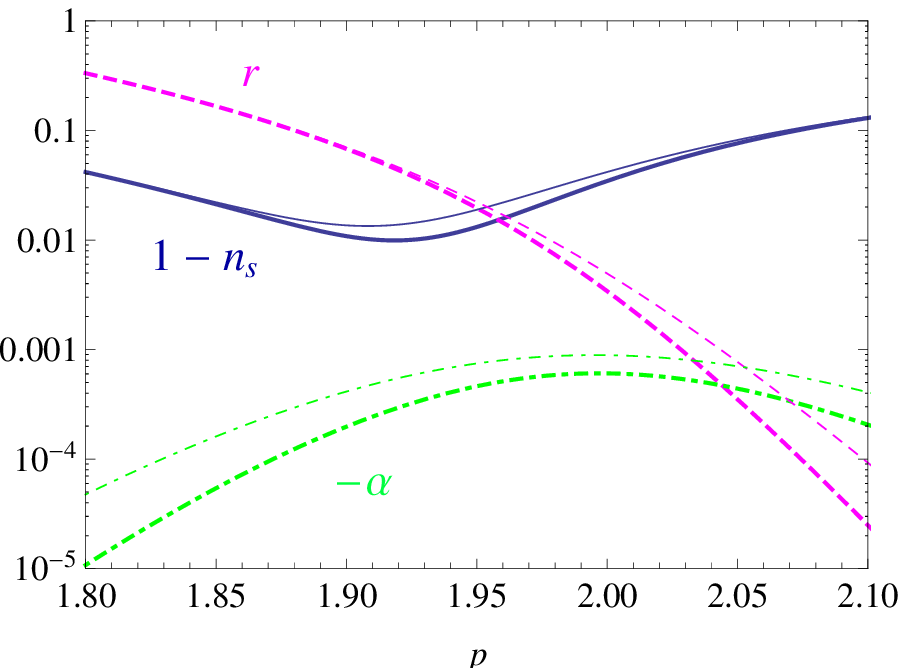}
\caption{
Left: Slow roll parameters, $\epsilon$ (blue solid), $|\eta|$ (magenta dashed), and $\xi$ (green dot-dashed). 
Right: Inflationary observables, $1-n_s$ (blue solid), $r$ (magenta dashed), and $-\alpha$ (green dot-dashed). The thick lines and thin lines are for $N_k=60$ and $50$, respectively. 
}
\label{fig:slow}
\end{figure}

In $R^2$ inflation, the hierarchy between the slow-roll parameters is $|\eta|\gg \epsilon\sim\xi$.
However, it is not the case for $R^p$ inflation with $p\not=2$.
The left panel of Fig.~\ref{fig:slow} exhibits the slow roll parameters \eqref{slowpn} for $p\approx 2$ with $N_k=60$ and $50$.
Blue solid, magenta dashed, and green dot-dashed lines are $\epsilon$, $|\eta|$, and $\xi$, respectively.
Thick lines are for $N_k=60$, while thin lines are for $N_k=50$.
Note that $\eta$ flips its sign at $p\simeq 1.94$ for $N_k=60$ ($p\simeq 1.93$ for $N=50$): $\eta>0$ for $p\lesssim 1.94$, and $\eta<0$ for $p\gtrsim 1.94$.
Now the hierarchy between the slow-roll parameters for $p\approx 2$ obviously varies from $|\eta| \gg \epsilon \sim \xi$ for $p= 2$.
However, we note that 
$\xi$ is always subleading. % for $1.8\leq p\leq 2.1$. 
Therefore, for the following, we treat $\epsilon$ and $\eta$ as the first order quantities, and $\xi$ as the second order quantity.

%==================== Figure ====================
%==================== Figure ====================

\section{Consistency relation}%%%%%%%%%%%%%%%%%%%%%%%%%%%%%%%%%%%%%%%%%
\label{sec-cr}

Now we want to relate the slow-roll parameters to the inflationary observables.
Since the comoving curvature perturbation and the tensor perturbation are invariant under the conformal transformation~\cite{Chiba:2008ia,Gong:2011qe}, we can make use of the slow-roll parameters obtained from the inflaton potential in the Einstein frame to evaluate the scalar spectral index $n_s$, its running $\alpha\equiv dn_s/d\ln k$, and the tensor-to-scalar ratio $r$.
Up to the leading order of the slow-roll parameters, the inflationary observables can be written as
\begin{align} 
n_s-1&=-6\epsilon+2\eta, \notag \\ 
r&=16\epsilon, \notag \\ 
\label{infobs} \alpha&=16\epsilon \eta-24 \epsilon^2-2\xi.
\end{align}
Let us remind that $\xi$ is treated as the second order quantity here. This treatment is valid for $R^p$ inflation and is also often implicitly assumed in the literature, but it is not necessarily always the case. 
For general case, where $\xi$ can be comparable to $\epsilon$ and $|\eta|$, we need more careful treatment~\cite{Motohashi:2015hpa}.

\subsection{$p=2$}%%%%%%%%%%%%%%%%%%%%%%%%%%%%%%%%%%%%%%%%%
\label{sec-con2}

For $p=2$, we can immediately write down \eqref{infobs} in terms of $N_k$ by the virtue of \eqref{slow2n}. Up to the leading order of $N_k^{-1}$, we obtain
\be 
\label{obs2} n_s-1=-\f{2}{N_k} , \quad 
r=\f{12}{N_k^2}, \quad
\alpha=-\f{2}{N_k^2} .
\ee
Thus the consistency relation is given by
\begin{align}
n_s-1=-\sqrt{\f{r}{3}},\quad
\alpha=-\f{r}{6}.
\end{align}
Equivalently, we can derive the above relation using  \eqref{conslow2} and \eqref{infobs}.

\subsection{$p\approx 2$}%%%%%%%%%%%%%%%%%%%%%%%%%%%%%%%%%%%%%%%%%
\label{sec-conp}

For general $p$, by substituting \eqref{slowpn} into \eqref{infobs}, we obtain 
\begin{align} 
n_s-1&= -\f{8(2-p)[ (2-p)E_k^2+p(E_k-1) ] }{3 [2 (p-1)E_k - p]^2} , \notag \\ 
r&=\f{64 E_k^2 (2-p)^2}{3 [2 (p-1)E_k - p]^2}, \notag \\
\label{alp} \alpha&= -\f{ 32p(2-p)^2 E_k(E_k-1) (2E_k-3p+4) }{9 [2 (p-1)E_k - p]^4} .
\end{align}
Thus, $n_s$, $r$, and $\alpha$ are related through the parameter $E_k= e^{4(2-p)N_k/(3p)}$. We can recover \eqref{obs2} if we take the limit $p\to 2$ in \eqref{alp}. By erasing $E_k$, we can obtain the consistency relation as
\begin{align} 
n_s-1&= -\f{(3p-4)}{\sqrt{3} p}\sqrt{r} - \f{3p-2}{8p}r + \f{8(2-p)}{3p} , \notag \\
\label{conp} \alpha&= \f{4(2-p)(3p-4)}{3\sqrt{3}p^2}\sqrt{r} - \f{15p^2-40p+24}{6p^2}r - \f{(3p-4)(4p-3)}{8\sqrt{3}p^2}r^{3/2} - \f{(p-1)(3p-2)}{32p^2}r^2.
\end{align}

In the right panel of Fig.~\ref{fig:slow}, we present the scalar spectral index, its running, and the tensor-to-scalar ratio for $p\approx 2$ with $N_k=60$ and $50$. Blue solid, magenta dashed, green dot-dashed lines are $(1-n_s)$, $r$, $-\alpha$, respectively, and thick and thin lines represent $N_k=60$ and $50$, respectively. 
We see that the scalar spectral index takes its maximum value $\simeq 0.99$ at $p\simeq 1.92$ and thus $R^p$ inflation describe only red-tilted spectrum. For $p<1.8$ or $p>2$, we have $n_s<0.96$.
On the other hand, the tensor-to-scalar ratio increases as $p$ decreases. Actually, $r$ exceeds $0.1$ and $0.2$ at $p\simeq 1.88$ and $p\simeq 1.84$, respectively, for $N_k=60$. 
As for the running of the scalar spectral index, $\alpha$ is always negative. Its amplitude takes the maximum value $\simeq 10^{-3}$ at $p\simeq 2$.

%==================== Figure ====================
\begin{figure}[t]
\centering
\includegraphics[width=78mm]{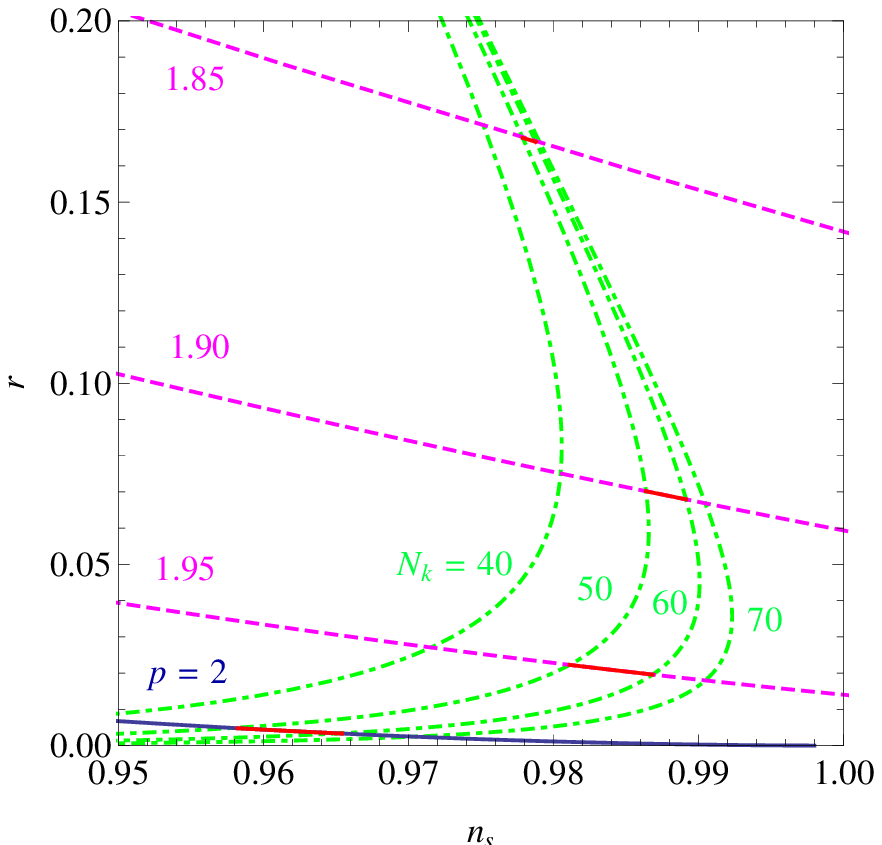}
\includegraphics[width=85mm]{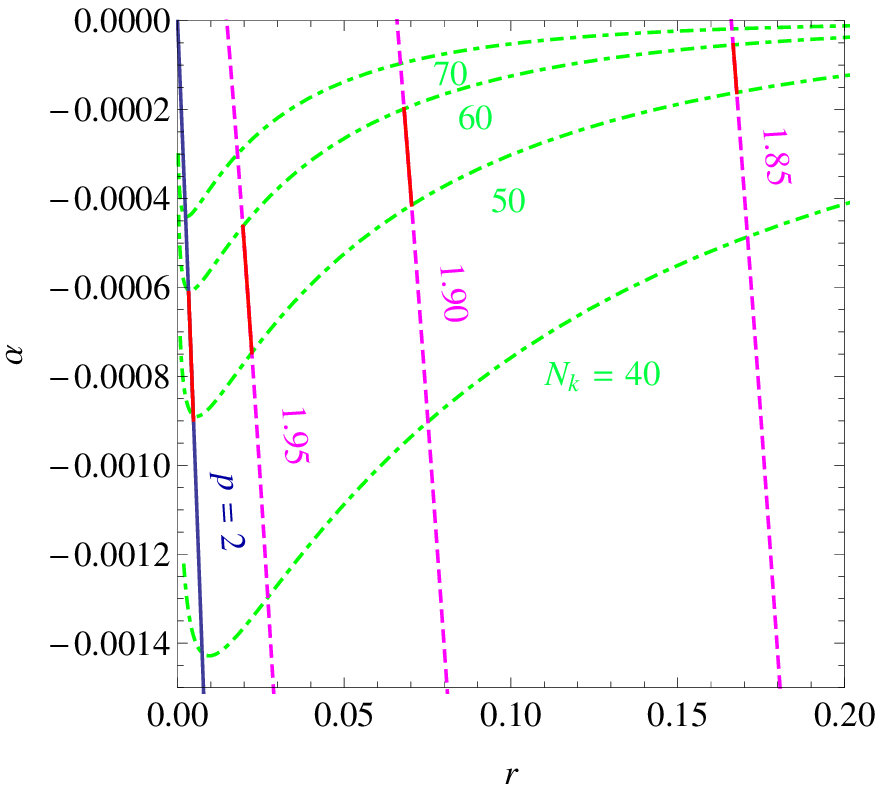}
\includegraphics[width=85mm]{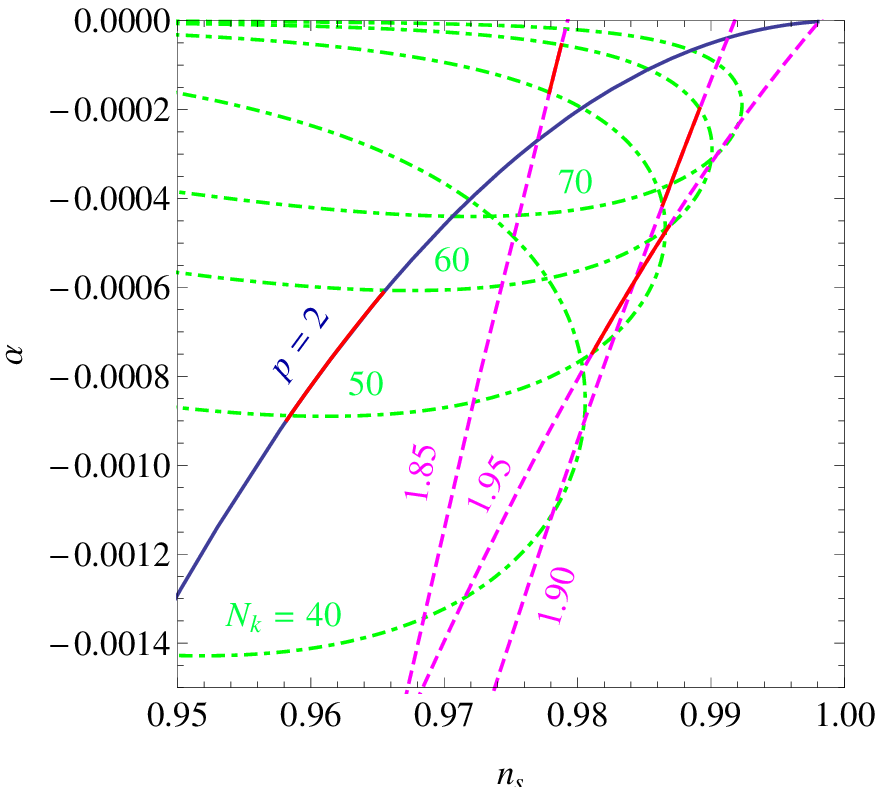}
\caption{
Scalar spectral index $n_s$, its running $\alpha$, and tensor-to-scalar ratio $r$ for $p=2$ (solid blue), and $1.95$, $1.90$, $1.85$ (magenta dashed), where e-folds between $N_k=50$ and $60$ are highlighted (red solid). Lines for fixed e-folds $N_k=40$, $50$, $60$, $70$ (green dot-dashed) are also shown.}
\label{fig:cr}
\end{figure}
%==================== Figure ====================

Using \eqref{alp} or \eqref{conp}, we can explicitly draw the consistency relation between the inflationary observables as presented in Fig.~\ref{fig:cr}.
Blue solid lines represent $p=2$, and magenta dashed lines represent $p=1.95$, $1.90$, $1.85$.  
We also show lines for fixed e-folds $N_k$ by green dot-dashed lines.
We highlighted lines for fixed $p$ with e-folds $50<N_k<60$. 
In particular, it is interesting that the scalar spectral index $n_s$ is sensitive for a deviation from $p=2$.
The panel for $(n_s,r)$ captures this property. 
For $1.8<p<2$, the spectral index varies as $0.96\lesssim n_s \lesssim 0.99$ but is always larger than $0.96$ for $N_k=60$. 
For $p \gtrsim 1.95$, the spectral index is very sensitive for $p$. Therefore, the parameter region $p \gtrsim 1.95$ is solely constrained by $n_s$.

We are also interested in how future constraint on $r$ tests the model.
From the panel for $(n_s,r)$ in Fig.~\ref{fig:cr}, we note that for $N_k=60$ small tensor-to-scalar ratio with $r \leq 0.05$ requires $1.92 \lesssim p \leq 2$ and $0.96 \lesssim n_s \lesssim 0.99$. 
For large $r$ with $0.05 \leq r \leq 0.1$, $p$ should be $1.88 \lesssim p\lesssim 1.92$ and $n_s$ needs to be within $0.98 \lesssim n_s \lesssim 0.99$. 
On the other hand, for fixed $n_s=0.96$, $r= 0.05$ and $0.1$ require  $(p,N_k)\simeq (1.93,30)$ and $(1.9,27)$, respectively.

From the panel for $(r,\alpha)$ in Fig.~\ref{fig:cr}, we can explicitly see that a deviation from $p=2$ suppresses $\alpha$, while $r$ is enhanced. This property is also useful to test $R^p$ inflation. We can constrain $p$ with an order $10^{-4}$ accuracy for $\alpha$.
The panel for $(n_s,\alpha)$ in Fig.~\ref{fig:cr} shows that it is difficult for this combination is to constrain $p$ because the lines are overlapping and thus there is a degeneracy between parameters. 
Therefore, in order to constrain $R^p$ inflation, it is important to measure both  
the scalar and the tensor spectra, namely, the combination of $(n_s,r)$ or $(r,\alpha)$ would constrain the model significantly.

\section{Conclusions}%%%%%%%%%%%%%%%%%%%%%%%%%%%%%%%%%%%%%%%%%
\label{sec-cn}

We investigated $R^p$ inflation with $p \approx 2$ in order to evaluate deviations from $R^2$ inflation. Using the inflaton potential in the Einstein frame, we explicitly wrote down the scalar spectral index $n_s$, its running $\alpha$, and the tensor-to-scalar ratio $r$ as in \eqref{alp}, which are presented in Fig.~\ref{fig:slow}.  
We can also explicitly draw the consistency relation as presented in Fig.~\ref{fig:cr}. We showed that the parameter region $p \gtrsim 1.95$ is solely constrained by $n_s$ and a precise measurement of $(n_s,r)$ or $(r,\alpha)$ can test a whole range of $p$.
Specifically, for $N_k=60$, $r \leq 0.05$ requires $1.92 \lesssim p \leq 2$ and $0.96 \lesssim n_s \lesssim 0.99$, while $0.05 \leq r \leq 0.1$ requires $1.88 \lesssim p\lesssim 1.92$ and $0.98 \lesssim n_s \lesssim 0.99$. 
On the other hand, for fixed $n_s=0.96$, $r\simeq 0.05$ and $0.1$ require  $(p,N_k)\simeq (1.93,30)$ and $(1.9,27)$, respectively.

\begin{acknowledgments}%%%%%%%%%%%%%%%%%%%%%%%%%%%%%%%%%%%%%%%%%
We thank W.\ Hu and and A.\ A.\ Starobinsky for useful discussions.
This work was supported by Japan Society for the Promotion of Science Postdoctoral Fellowships for Research Abroad.
\end{acknowledgments}

\bibliography{refs}

%merlin.mbs apsrev4-1.bst 2010-07-25 4.21a (PWD, AO, DPC) hacked
%Control: key (0)
%Control: author (8) initials jnrlst
%Control: editor formatted (1) identically to author
%Control: production of article title (-1) disabled
%Control: page (0) single
%Control: year (1) truncated
%Control: production of eprint (0) enabled
\begin{thebibliography}{43}%
\makeatletter
\providecommand \@ifxundefined [1]{%
 \@ifx{#1\undefined}
}%
\providecommand \@ifnum [1]{%
 \ifnum #1\expandafter \@firstoftwo
 \else \expandafter \@secondoftwo
 \fi
}%
\providecommand \@ifx [1]{%
 \ifx #1\expandafter \@firstoftwo
 \else \expandafter \@secondoftwo
 \fi
}%
\providecommand \natexlab [1]{#1}%
\providecommand \enquote  [1]{``#1''}%
\providecommand \bibnamefont  [1]{#1}%
\providecommand \bibfnamefont [1]{#1}%
\providecommand \citenamefont [1]{#1}%
\providecommand \href@noop [0]{\@secondoftwo}%
\providecommand \href [0]{\begingroup \@sanitize@url \@href}%
\providecommand \@href[1]{\@@startlink{#1}\@@href}%
\providecommand \@@href[1]{\endgroup#1\@@endlink}%
\providecommand \@sanitize@url [0]{\catcode `\\12\catcode `\$12\catcode
  `\&12\catcode `\#12\catcode `\^12\catcode `\_12\catcode `\%12\relax}%
\providecommand \@@startlink[1]{}%
\providecommand \@@endlink[0]{}%
\providecommand \url  [0]{\begingroup\@sanitize@url \@url }%
\providecommand \@url [1]{\endgroup\@href {#1}{\urlprefix }}%
\providecommand \urlprefix  [0]{URL }%
\providecommand \Eprint [0]{\href }%
\providecommand \doibase [0]{http://dx.doi.org/}%
\providecommand \selectlanguage [0]{\@gobble}%
\providecommand \bibinfo  [0]{\@secondoftwo}%
\providecommand \bibfield  [0]{\@secondoftwo}%
\providecommand \translation [1]{[#1]}%
\providecommand \BibitemOpen [0]{}%
\providecommand \bibitemStop [0]{}%
\providecommand \bibitemNoStop [0]{.\EOS\space}%
\providecommand \EOS [0]{\spacefactor3000\relax}%
\providecommand \BibitemShut  [1]{\csname bibitem#1\endcsname}%
\let\auto@bib@innerbib\@empty
%</preamble>
\bibitem [{\citenamefont {Starobinsky}(1980)}]{Starobinsky:1980te}%
  \BibitemOpen
  \bibfield  {author} {\bibinfo {author} {\bibfnamefont {A.~A.}\ \bibnamefont
  {Starobinsky}},\ }\href {\doibase 10.1016/0370-2693(80)90670-X} {\bibfield
  {journal} {\bibinfo  {journal} {Phys.Lett.}\ }\textbf {\bibinfo {volume}
  {B91}},\ \bibinfo {pages} {99} (\bibinfo {year} {1980})}\BibitemShut
  {NoStop}%
%%CITATION = PHLTA,B91,99;%%
\bibitem [{\citenamefont {Vilenkin}(1985)}]{Vilenkin:1985md}%
  \BibitemOpen
  \bibfield  {author} {\bibinfo {author} {\bibfnamefont {A.}~\bibnamefont
  {Vilenkin}},\ }\href {\doibase 10.1103/PhysRevD.32.2511} {\bibfield
  {journal} {\bibinfo  {journal} {Phys.Rev.}\ }\textbf {\bibinfo {volume}
  {D32}},\ \bibinfo {pages} {2511} (\bibinfo {year} {1985})}\BibitemShut
  {NoStop}%
%%CITATION = PHRVA,D32,2511;%%
\bibitem [{\citenamefont {Mijic}\ \emph {et~al.}(1986)\citenamefont {Mijic},
  \citenamefont {Morris},\ and\ \citenamefont {Suen}}]{Mijic:1986iv}%
  \BibitemOpen
  \bibfield  {author} {\bibinfo {author} {\bibfnamefont {M.~B.}\ \bibnamefont
  {Mijic}}, \bibinfo {author} {\bibfnamefont {M.~S.}\ \bibnamefont {Morris}}, \
  and\ \bibinfo {author} {\bibfnamefont {W.-M.}\ \bibnamefont {Suen}},\ }\href
  {\doibase 10.1103/PhysRevD.34.2934} {\bibfield  {journal} {\bibinfo
  {journal} {Phys.Rev.}\ }\textbf {\bibinfo {volume} {D34}},\ \bibinfo {pages}
  {2934} (\bibinfo {year} {1986})}\BibitemShut {NoStop}%
%%CITATION = PHRVA,D34,2934;%%
\bibitem [{\citenamefont {Ford}(1987)}]{Ford:1986sy}%
  \BibitemOpen
  \bibfield  {author} {\bibinfo {author} {\bibfnamefont {L.}~\bibnamefont
  {Ford}},\ }\href {\doibase 10.1103/PhysRevD.35.2955} {\bibfield  {journal}
  {\bibinfo  {journal} {Phys.Rev.}\ }\textbf {\bibinfo {volume} {D35}},\
  \bibinfo {pages} {2955} (\bibinfo {year} {1987})}\BibitemShut {NoStop}%
%%CITATION = PHRVA,D35,2955;%%
\bibitem [{\citenamefont {Hinshaw}\ \emph {et~al.}(2013)\citenamefont {Hinshaw}
  \emph {et~al.}}]{Hinshaw:2012aka}%
  \BibitemOpen
  \bibfield  {author} {\bibinfo {author} {\bibfnamefont {G.}~\bibnamefont
  {Hinshaw}} \emph {et~al.} (\bibinfo {collaboration} {WMAP}),\ }\href
  {\doibase 10.1088/0067-0049/208/2/19} {\bibfield  {journal} {\bibinfo
  {journal} {Astrophys.J.Suppl.}\ }\textbf {\bibinfo {volume} {208}},\ \bibinfo
  {pages} {19} (\bibinfo {year} {2013})},\ \Eprint
  {http://arxiv.org/abs/1212.5226} {arXiv:1212.5226 [astro-ph.CO]} \BibitemShut
  {NoStop}%
%%CITATION = ARXIV:1212.5226;%%
\bibitem [{\citenamefont {Ade}\ \emph {et~al.}(2013)\citenamefont {Ade} \emph
  {et~al.}}]{Ade:2013uln}%
  \BibitemOpen
  \bibfield  {author} {\bibinfo {author} {\bibfnamefont {P.}~\bibnamefont
  {Ade}} \emph {et~al.} (\bibinfo {collaboration} {Planck Collaboration}),\
  }\href@noop {} {\  (\bibinfo {year} {2013})},\ \Eprint
  {http://arxiv.org/abs/1303.5082} {arXiv:1303.5082 [astro-ph.CO]} \BibitemShut
  {NoStop}%
%%CITATION = ARXIV:1303.5082;%%
\bibitem [{\citenamefont {Ade}\ \emph {et~al.}(2014)\citenamefont {Ade} \emph
  {et~al.}}]{Ade:2014xna}%
  \BibitemOpen
  \bibfield  {author} {\bibinfo {author} {\bibfnamefont {P.}~\bibnamefont
  {Ade}} \emph {et~al.} (\bibinfo {collaboration} {BICEP2 Collaboration}),\
  }\href {\doibase 10.1103/PhysRevLett.112.241101} {\bibfield  {journal}
  {\bibinfo  {journal} {Phys.Rev.Lett.}\ }\textbf {\bibinfo {volume} {112}},\
  \bibinfo {pages} {241101} (\bibinfo {year} {2014})},\ \Eprint
  {http://arxiv.org/abs/1403.3985} {arXiv:1403.3985 [astro-ph.CO]} \BibitemShut
  {NoStop}%
%%CITATION = ARXIV:1403.3985;%%
\bibitem [{\citenamefont {Adam}\ \emph {et~al.}(2014)\citenamefont {Adam} \emph
  {et~al.}}]{Adam:2014bub}%
  \BibitemOpen
  \bibfield  {author} {\bibinfo {author} {\bibfnamefont {R.}~\bibnamefont
  {Adam}} \emph {et~al.} (\bibinfo {collaboration} {Planck Collaboration}),\
  }\href@noop {} {\  (\bibinfo {year} {2014})},\ \Eprint
  {http://arxiv.org/abs/1409.5738} {arXiv:1409.5738 [astro-ph.CO]} \BibitemShut
  {NoStop}%
%%CITATION = ARXIV:1409.5738;%%
\bibitem [{\citenamefont {Hu}\ and\ \citenamefont {Sawicki}(2007)}]{Hu:2007nk}%
  \BibitemOpen
  \bibfield  {author} {\bibinfo {author} {\bibfnamefont {W.}~\bibnamefont
  {Hu}}\ and\ \bibinfo {author} {\bibfnamefont {I.}~\bibnamefont {Sawicki}},\
  }\href {\doibase 10.1103/PhysRevD.76.064004} {\bibfield  {journal} {\bibinfo
  {journal} {Phys.Rev.}\ }\textbf {\bibinfo {volume} {D76}},\ \bibinfo {pages}
  {064004} (\bibinfo {year} {2007})},\ \Eprint {http://arxiv.org/abs/0705.1158}
  {arXiv:0705.1158 [astro-ph]} \BibitemShut {NoStop}%
%%CITATION = ARXIV:0705.1158;%%
\bibitem [{\citenamefont {Appleby}\ and\ \citenamefont
  {Battye}(2007)}]{Appleby:2007vb}%
  \BibitemOpen
  \bibfield  {author} {\bibinfo {author} {\bibfnamefont {S.~A.}\ \bibnamefont
  {Appleby}}\ and\ \bibinfo {author} {\bibfnamefont {R.~A.}\ \bibnamefont
  {Battye}},\ }\href {\doibase 10.1016/j.physletb.2007.08.037} {\bibfield
  {journal} {\bibinfo  {journal} {Phys.Lett.}\ }\textbf {\bibinfo {volume}
  {B654}},\ \bibinfo {pages} {7} (\bibinfo {year} {2007})},\ \Eprint
  {http://arxiv.org/abs/0705.3199} {arXiv:0705.3199 [astro-ph]} \BibitemShut
  {NoStop}%
%%CITATION = ARXIV:0705.3199;%%
\bibitem [{\citenamefont {Starobinsky}(2007)}]{Starobinsky:2007hu}%
  \BibitemOpen
  \bibfield  {author} {\bibinfo {author} {\bibfnamefont {A.~A.}\ \bibnamefont
  {Starobinsky}},\ }\href {\doibase 10.1134/S0021364007150027} {\bibfield
  {journal} {\bibinfo  {journal} {JETP Lett.}\ }\textbf {\bibinfo {volume}
  {86}},\ \bibinfo {pages} {157} (\bibinfo {year} {2007})},\ \Eprint
  {http://arxiv.org/abs/0706.2041} {arXiv:0706.2041 [astro-ph]} \BibitemShut
  {NoStop}%
%%CITATION = ARXIV:0706.2041;%%
\bibitem [{\citenamefont {Motohashi}\ \emph
  {et~al.}(2010{\natexlab{a}})\citenamefont {Motohashi}, \citenamefont
  {Starobinsky},\ and\ \citenamefont {Yokoyama}}]{Motohashi:2010tb}%
  \BibitemOpen
  \bibfield  {author} {\bibinfo {author} {\bibfnamefont {H.}~\bibnamefont
  {Motohashi}}, \bibinfo {author} {\bibfnamefont {A.~A.}\ \bibnamefont
  {Starobinsky}}, \ and\ \bibinfo {author} {\bibfnamefont {J.}~\bibnamefont
  {Yokoyama}},\ }\href {\doibase 10.1143/PTP.123.887} {\bibfield  {journal}
  {\bibinfo  {journal} {Prog.Theor.Phys.}\ }\textbf {\bibinfo {volume} {123}},\
  \bibinfo {pages} {887} (\bibinfo {year} {2010}{\natexlab{a}})},\ \Eprint
  {http://arxiv.org/abs/1002.1141} {arXiv:1002.1141 [astro-ph.CO]} \BibitemShut
  {NoStop}%
%%CITATION = ARXIV:1002.1141;%%
\bibitem [{\citenamefont {Motohashi}\ \emph {et~al.}(2011)\citenamefont
  {Motohashi}, \citenamefont {Starobinsky},\ and\ \citenamefont
  {Yokoyama}}]{Motohashi:2011wy}%
  \BibitemOpen
  \bibfield  {author} {\bibinfo {author} {\bibfnamefont {H.}~\bibnamefont
  {Motohashi}}, \bibinfo {author} {\bibfnamefont {A.~A.}\ \bibnamefont
  {Starobinsky}}, \ and\ \bibinfo {author} {\bibfnamefont {J.}~\bibnamefont
  {Yokoyama}},\ }\href {\doibase 10.1088/1475-7516/2011/06/006} {\bibfield
  {journal} {\bibinfo  {journal} {JCAP}\ }\textbf {\bibinfo {volume} {1106}},\
  \bibinfo {pages} {006} (\bibinfo {year} {2011})},\ \Eprint
  {http://arxiv.org/abs/1101.0744} {arXiv:1101.0744 [astro-ph.CO]} \BibitemShut
  {NoStop}%
%%CITATION = ARXIV:1101.0744;%%
\bibitem [{\citenamefont {Gannouji}\ \emph {et~al.}(2009)\citenamefont
  {Gannouji}, \citenamefont {Moraes},\ and\ \citenamefont
  {Polarski}}]{Gannouji:2008wt}%
  \BibitemOpen
  \bibfield  {author} {\bibinfo {author} {\bibfnamefont {R.}~\bibnamefont
  {Gannouji}}, \bibinfo {author} {\bibfnamefont {B.}~\bibnamefont {Moraes}}, \
  and\ \bibinfo {author} {\bibfnamefont {D.}~\bibnamefont {Polarski}},\ }\href
  {\doibase 10.1088/1475-7516/2009/02/034} {\bibfield  {journal} {\bibinfo
  {journal} {JCAP}\ }\textbf {\bibinfo {volume} {0902}},\ \bibinfo {pages}
  {034} (\bibinfo {year} {2009})},\ \Eprint {http://arxiv.org/abs/0809.3374}
  {arXiv:0809.3374 [astro-ph]} \BibitemShut {NoStop}%
%%CITATION = ARXIV:0809.3374;%%
\bibitem [{\citenamefont {Motohashi}\ \emph {et~al.}(2009)\citenamefont
  {Motohashi}, \citenamefont {Starobinsky},\ and\ \citenamefont
  {Yokoyama}}]{Motohashi:2009qn}%
  \BibitemOpen
  \bibfield  {author} {\bibinfo {author} {\bibfnamefont {H.}~\bibnamefont
  {Motohashi}}, \bibinfo {author} {\bibfnamefont {A.~A.}\ \bibnamefont
  {Starobinsky}}, \ and\ \bibinfo {author} {\bibfnamefont {J.}~\bibnamefont
  {Yokoyama}},\ }\href {\doibase 10.1142/S0218271809015278} {\bibfield
  {journal} {\bibinfo  {journal} {Int.J.Mod.Phys.}\ }\textbf {\bibinfo {volume}
  {D18}},\ \bibinfo {pages} {1731} (\bibinfo {year} {2009})},\ \Eprint
  {http://arxiv.org/abs/0905.0730} {arXiv:0905.0730 [astro-ph.CO]} \BibitemShut
  {NoStop}%
%%CITATION = ARXIV:0905.0730;%%
\bibitem [{\citenamefont {Tsujikawa}\ \emph {et~al.}(2009)\citenamefont
  {Tsujikawa}, \citenamefont {Gannouji}, \citenamefont {Moraes},\ and\
  \citenamefont {Polarski}}]{Tsujikawa:2009ku}%
  \BibitemOpen
  \bibfield  {author} {\bibinfo {author} {\bibfnamefont {S.}~\bibnamefont
  {Tsujikawa}}, \bibinfo {author} {\bibfnamefont {R.}~\bibnamefont {Gannouji}},
  \bibinfo {author} {\bibfnamefont {B.}~\bibnamefont {Moraes}}, \ and\ \bibinfo
  {author} {\bibfnamefont {D.}~\bibnamefont {Polarski}},\ }\href {\doibase
  10.1103/PhysRevD.80.084044} {\bibfield  {journal} {\bibinfo  {journal}
  {Phys.Rev.}\ }\textbf {\bibinfo {volume} {D80}},\ \bibinfo {pages} {084044}
  (\bibinfo {year} {2009})},\ \Eprint {http://arxiv.org/abs/0908.2669}
  {arXiv:0908.2669 [astro-ph.CO]} \BibitemShut {NoStop}%
%%CITATION = ARXIV:0908.2669;%%
\bibitem [{\citenamefont {Motohashi}\ \emph
  {et~al.}(2010{\natexlab{b}})\citenamefont {Motohashi}, \citenamefont
  {Starobinsky},\ and\ \citenamefont {Yokoyama}}]{Motohashi:2010sj}%
  \BibitemOpen
  \bibfield  {author} {\bibinfo {author} {\bibfnamefont {H.}~\bibnamefont
  {Motohashi}}, \bibinfo {author} {\bibfnamefont {A.~A.}\ \bibnamefont
  {Starobinsky}}, \ and\ \bibinfo {author} {\bibfnamefont {J.}~\bibnamefont
  {Yokoyama}},\ }\href {\doibase 10.1143/PTP.124.541} {\bibfield  {journal}
  {\bibinfo  {journal} {Prog.Theor.Phys.}\ }\textbf {\bibinfo {volume} {124}},\
  \bibinfo {pages} {541} (\bibinfo {year} {2010}{\natexlab{b}})},\ \Eprint
  {http://arxiv.org/abs/1005.1171} {arXiv:1005.1171 [astro-ph.CO]} \BibitemShut
  {NoStop}%
%%CITATION = ARXIV:1005.1171;%%
\bibitem [{\citenamefont {Motohashi}\ \emph {et~al.}(2013)\citenamefont
  {Motohashi}, \citenamefont {Starobinsky},\ and\ \citenamefont
  {Yokoyama}}]{Motohashi:2012wc}%
  \BibitemOpen
  \bibfield  {author} {\bibinfo {author} {\bibfnamefont {H.}~\bibnamefont
  {Motohashi}}, \bibinfo {author} {\bibfnamefont {A.~A.}\ \bibnamefont
  {Starobinsky}}, \ and\ \bibinfo {author} {\bibfnamefont {J.}~\bibnamefont
  {Yokoyama}},\ }\href {\doibase 10.1103/PhysRevLett.110.121302} {\bibfield
  {journal} {\bibinfo  {journal} {Phys.Rev.Lett.}\ }\textbf {\bibinfo {volume}
  {110}},\ \bibinfo {pages} {121302} (\bibinfo {year} {2013})},\ \Eprint
  {http://arxiv.org/abs/1203.6828} {arXiv:1203.6828 [astro-ph.CO]} \BibitemShut
  {NoStop}%
%%CITATION = ARXIV:1203.6828;%%
\bibitem [{\citenamefont {Tsujikawa}(2008)}]{Tsujikawa:2007xu}%
  \BibitemOpen
  \bibfield  {author} {\bibinfo {author} {\bibfnamefont {S.}~\bibnamefont
  {Tsujikawa}},\ }\href {\doibase 10.1103/PhysRevD.77.023507} {\bibfield
  {journal} {\bibinfo  {journal} {Phys.Rev.}\ }\textbf {\bibinfo {volume}
  {D77}},\ \bibinfo {pages} {023507} (\bibinfo {year} {2008})},\ \Eprint
  {http://arxiv.org/abs/0709.1391} {arXiv:0709.1391 [astro-ph]} \BibitemShut
  {NoStop}%
%%CITATION = ARXIV:0709.1391;%%
\bibitem [{\citenamefont {Appleby}\ and\ \citenamefont
  {Battye}(2008)}]{Appleby:2008tv}%
  \BibitemOpen
  \bibfield  {author} {\bibinfo {author} {\bibfnamefont {S.}~\bibnamefont
  {Appleby}}\ and\ \bibinfo {author} {\bibfnamefont {R.}~\bibnamefont
  {Battye}},\ }\href {\doibase 10.1088/1475-7516/2008/05/019} {\bibfield
  {journal} {\bibinfo  {journal} {JCAP}\ }\textbf {\bibinfo {volume} {0805}},\
  \bibinfo {pages} {019} (\bibinfo {year} {2008})},\ \Eprint
  {http://arxiv.org/abs/0803.1081} {arXiv:0803.1081 [astro-ph]} \BibitemShut
  {NoStop}%
%%CITATION = ARXIV:0803.1081;%%
\bibitem [{\citenamefont {Frolov}(2008)}]{Frolov:2008uf}%
  \BibitemOpen
  \bibfield  {author} {\bibinfo {author} {\bibfnamefont {A.~V.}\ \bibnamefont
  {Frolov}},\ }\href {\doibase 10.1103/PhysRevLett.101.061103} {\bibfield
  {journal} {\bibinfo  {journal} {Phys.Rev.Lett.}\ }\textbf {\bibinfo {volume}
  {101}},\ \bibinfo {pages} {061103} (\bibinfo {year} {2008})},\ \Eprint
  {http://arxiv.org/abs/0803.2500} {arXiv:0803.2500 [astro-ph]} \BibitemShut
  {NoStop}%
%%CITATION = ARXIV:0803.2500;%%
\bibitem [{\citenamefont {Kobayashi}\ and\ \citenamefont
  {Maeda}(2008)}]{Kobayashi:2008tq}%
  \BibitemOpen
  \bibfield  {author} {\bibinfo {author} {\bibfnamefont {T.}~\bibnamefont
  {Kobayashi}}\ and\ \bibinfo {author} {\bibfnamefont {K.-i.}\ \bibnamefont
  {Maeda}},\ }\href {\doibase 10.1103/PhysRevD.78.064019} {\bibfield  {journal}
  {\bibinfo  {journal} {Phys.Rev.}\ }\textbf {\bibinfo {volume} {D78}},\
  \bibinfo {pages} {064019} (\bibinfo {year} {2008})},\ \Eprint
  {http://arxiv.org/abs/0807.2503} {arXiv:0807.2503 [astro-ph]} \BibitemShut
  {NoStop}%
%%CITATION = ARXIV:0807.2503;%%
\bibitem [{\citenamefont {Appleby}\ \emph {et~al.}(2010)\citenamefont
  {Appleby}, \citenamefont {Battye},\ and\ \citenamefont
  {Starobinsky}}]{Appleby:2009uf}%
  \BibitemOpen
  \bibfield  {author} {\bibinfo {author} {\bibfnamefont {S.~A.}\ \bibnamefont
  {Appleby}}, \bibinfo {author} {\bibfnamefont {R.~A.}\ \bibnamefont {Battye}},
  \ and\ \bibinfo {author} {\bibfnamefont {A.~A.}\ \bibnamefont
  {Starobinsky}},\ }\href {\doibase 10.1088/1475-7516/2010/06/005} {\bibfield
  {journal} {\bibinfo  {journal} {JCAP}\ }\textbf {\bibinfo {volume} {1006}},\
  \bibinfo {pages} {005} (\bibinfo {year} {2010})},\ \Eprint
  {http://arxiv.org/abs/0909.1737} {arXiv:0909.1737 [astro-ph.CO]} \BibitemShut
  {NoStop}%
%%CITATION = ARXIV:0909.1737;%%
\bibitem [{\citenamefont {Motohashi}\ and\ \citenamefont
  {Nishizawa}(2012)}]{Motohashi:2012tt}%
  \BibitemOpen
  \bibfield  {author} {\bibinfo {author} {\bibfnamefont {H.}~\bibnamefont
  {Motohashi}}\ and\ \bibinfo {author} {\bibfnamefont {A.}~\bibnamefont
  {Nishizawa}},\ }\href {\doibase 10.1103/PhysRevD.86.083514} {\bibfield
  {journal} {\bibinfo  {journal} {Phys.Rev.}\ }\textbf {\bibinfo {volume}
  {D86}},\ \bibinfo {pages} {083514} (\bibinfo {year} {2012})},\ \Eprint
  {http://arxiv.org/abs/1204.1472} {arXiv:1204.1472 [astro-ph.CO]} \BibitemShut
  {NoStop}%
%%CITATION = ARXIV:1204.1472;%%
\bibitem [{\citenamefont {Nishizawa}\ and\ \citenamefont
  {Motohashi}(2014)}]{Nishizawa:2014zra}%
  \BibitemOpen
  \bibfield  {author} {\bibinfo {author} {\bibfnamefont {A.}~\bibnamefont
  {Nishizawa}}\ and\ \bibinfo {author} {\bibfnamefont {H.}~\bibnamefont
  {Motohashi}},\ }\href {\doibase 10.1103/PhysRevD.89.063541} {\bibfield
  {journal} {\bibinfo  {journal} {Phys.Rev.}\ }\textbf {\bibinfo {volume}
  {D89}},\ \bibinfo {pages} {063541} (\bibinfo {year} {2014})},\ \Eprint
  {http://arxiv.org/abs/1401.1023} {arXiv:1401.1023 [astro-ph.CO]} \BibitemShut
  {NoStop}%
%%CITATION = ARXIV:1401.1023;%%
\bibitem [{\citenamefont {Schmidt}(1989)}]{Schmidt:1989zz}%
  \BibitemOpen
  \bibfield  {author} {\bibinfo {author} {\bibfnamefont {H.}~\bibnamefont
  {Schmidt}},\ }\href {\doibase 10.1088/0264-9381/6/4/013} {\bibfield
  {journal} {\bibinfo  {journal} {Class.Quant.Grav.}\ }\textbf {\bibinfo
  {volume} {6}},\ \bibinfo {pages} {557} (\bibinfo {year} {1989})}\BibitemShut
  {NoStop}%
%%CITATION = CQGRD,6,557;%%
\bibitem [{\citenamefont {Maeda}(1989)}]{Maeda:1988ab}%
  \BibitemOpen
  \bibfield  {author} {\bibinfo {author} {\bibfnamefont {K.-i.}\ \bibnamefont
  {Maeda}},\ }\href {\doibase 10.1103/PhysRevD.39.3159} {\bibfield  {journal}
  {\bibinfo  {journal} {Phys.Rev.}\ }\textbf {\bibinfo {volume} {D39}},\
  \bibinfo {pages} {3159} (\bibinfo {year} {1989})}\BibitemShut {NoStop}%
%%CITATION = PHRVA,D39,3159;%%
\bibitem [{\citenamefont {Muller}\ \emph {et~al.}(1990)\citenamefont {Muller},
  \citenamefont {Schmidt},\ and\ \citenamefont {Starobinsky}}]{Muller:1989rp}%
  \BibitemOpen
  \bibfield  {author} {\bibinfo {author} {\bibfnamefont {V.}~\bibnamefont
  {Muller}}, \bibinfo {author} {\bibfnamefont {H.}~\bibnamefont {Schmidt}}, \
  and\ \bibinfo {author} {\bibfnamefont {A.~A.}\ \bibnamefont {Starobinsky}},\
  }\href {\doibase 10.1088/0264-9381/7/7/012} {\bibfield  {journal} {\bibinfo
  {journal} {Class.Quant.Grav.}\ }\textbf {\bibinfo {volume} {7}},\ \bibinfo
  {pages} {1163} (\bibinfo {year} {1990})}\BibitemShut {NoStop}%
%%CITATION = CQGRD,7,1163;%%
\bibitem [{\citenamefont {Gottlober}\ \emph {et~al.}(1992)\citenamefont
  {Gottlober}, \citenamefont {Muller}, \citenamefont {Schmidt},\ and\
  \citenamefont {Starobinsky}}]{Gottlober:1992rg}%
  \BibitemOpen
  \bibfield  {author} {\bibinfo {author} {\bibfnamefont {S.}~\bibnamefont
  {Gottlober}}, \bibinfo {author} {\bibfnamefont {V.}~\bibnamefont {Muller}},
  \bibinfo {author} {\bibfnamefont {H.}~\bibnamefont {Schmidt}}, \ and\
  \bibinfo {author} {\bibfnamefont {A.~A.}\ \bibnamefont {Starobinsky}},\
  }\href {\doibase 10.1142/S0218271892000136} {\bibfield  {journal} {\bibinfo
  {journal} {Int.J.Mod.Phys.}\ }\textbf {\bibinfo {volume} {D1}},\ \bibinfo
  {pages} {257} (\bibinfo {year} {1992})}\BibitemShut {NoStop}%
%%CITATION = IMPAE,D1,257;%%
\bibitem [{\citenamefont {De~Felice}\ and\ \citenamefont
  {Tsujikawa}(2010)}]{DeFelice:2010aj}%
  \BibitemOpen
  \bibfield  {author} {\bibinfo {author} {\bibfnamefont {A.}~\bibnamefont
  {De~Felice}}\ and\ \bibinfo {author} {\bibfnamefont {S.}~\bibnamefont
  {Tsujikawa}},\ }\href {\doibase 10.12942/lrr-2010-3} {\bibfield  {journal}
  {\bibinfo  {journal} {Living Rev.Rel.}\ }\textbf {\bibinfo {volume} {13}},\
  \bibinfo {pages} {3} (\bibinfo {year} {2010})},\ \Eprint
  {http://arxiv.org/abs/1002.4928} {arXiv:1002.4928 [gr-qc]} \BibitemShut
  {NoStop}%
%%CITATION = ARXIV:1002.4928;%%
\bibitem [{\citenamefont {Martin}\ \emph
  {et~al.}(2014{\natexlab{a}})\citenamefont {Martin}, \citenamefont
  {Ringeval},\ and\ \citenamefont {Vennin}}]{Martin:2014vha}%
  \BibitemOpen
  \bibfield  {author} {\bibinfo {author} {\bibfnamefont {J.}~\bibnamefont
  {Martin}}, \bibinfo {author} {\bibfnamefont {C.}~\bibnamefont {Ringeval}}, \
  and\ \bibinfo {author} {\bibfnamefont {V.}~\bibnamefont {Vennin}},\ }\href
  {\doibase 10.1016/j.dark.2014.01.003} {\bibfield  {journal} {\bibinfo
  {journal} {Phys.Dark Univ.}\ } (\bibinfo {year} {2014}{\natexlab{a}}),\
  10.1016/j.dark.2014.01.003},\ \Eprint {http://arxiv.org/abs/1303.3787}
  {arXiv:1303.3787 [astro-ph.CO]} \BibitemShut {NoStop}%
%%CITATION = ARXIV:1303.3787;%%
\bibitem [{\citenamefont {Martin}\ \emph
  {et~al.}(2014{\natexlab{b}})\citenamefont {Martin}, \citenamefont {Ringeval},
  \citenamefont {Trotta},\ and\ \citenamefont {Vennin}}]{Martin:2013nzq}%
  \BibitemOpen
  \bibfield  {author} {\bibinfo {author} {\bibfnamefont {J.}~\bibnamefont
  {Martin}}, \bibinfo {author} {\bibfnamefont {C.}~\bibnamefont {Ringeval}},
  \bibinfo {author} {\bibfnamefont {R.}~\bibnamefont {Trotta}}, \ and\ \bibinfo
  {author} {\bibfnamefont {V.}~\bibnamefont {Vennin}},\ }\href {\doibase
  10.1088/1475-7516/2014/03/039} {\bibfield  {journal} {\bibinfo  {journal}
  {JCAP}\ }\textbf {\bibinfo {volume} {1403}},\ \bibinfo {pages} {039}
  (\bibinfo {year} {2014}{\natexlab{b}})},\ \Eprint
  {http://arxiv.org/abs/1312.3529} {arXiv:1312.3529 [astro-ph.CO]} \BibitemShut
  {NoStop}%
%%CITATION = ARXIV:1312.3529;%%
\bibitem [{\citenamefont {Codello}\ \emph {et~al.}(2015)\citenamefont
  {Codello}, \citenamefont {Joergensen}, \citenamefont {Sannino},\ and\
  \citenamefont {Svendsen}}]{Codello:2014sua}%
  \BibitemOpen
  \bibfield  {author} {\bibinfo {author} {\bibfnamefont {A.}~\bibnamefont
  {Codello}}, \bibinfo {author} {\bibfnamefont {J.}~\bibnamefont {Joergensen}},
  \bibinfo {author} {\bibfnamefont {F.}~\bibnamefont {Sannino}}, \ and\
  \bibinfo {author} {\bibfnamefont {O.}~\bibnamefont {Svendsen}},\ }\href
  {\doibase 10.1007/JHEP02(2015)050} {\bibfield  {journal} {\bibinfo  {journal}
  {JHEP}\ }\textbf {\bibinfo {volume} {1502}},\ \bibinfo {pages} {050}
  (\bibinfo {year} {2015})},\ \Eprint {http://arxiv.org/abs/1404.3558}
  {arXiv:1404.3558 [hep-ph]} \BibitemShut {NoStop}%
%%CITATION = ARXIV:1404.3558;%%
\bibitem [{\citenamefont {Martin}\ \emph
  {et~al.}(2014{\natexlab{c}})\citenamefont {Martin}, \citenamefont {Ringeval},
  \citenamefont {Trotta},\ and\ \citenamefont {Vennin}}]{Martin:2014lra}%
  \BibitemOpen
  \bibfield  {author} {\bibinfo {author} {\bibfnamefont {J.}~\bibnamefont
  {Martin}}, \bibinfo {author} {\bibfnamefont {C.}~\bibnamefont {Ringeval}},
  \bibinfo {author} {\bibfnamefont {R.}~\bibnamefont {Trotta}}, \ and\ \bibinfo
  {author} {\bibfnamefont {V.}~\bibnamefont {Vennin}},\ }\href {\doibase
  10.1103/PhysRevD.90.063501} {\bibfield  {journal} {\bibinfo  {journal}
  {Phys.Rev.}\ }\textbf {\bibinfo {volume} {D90}},\ \bibinfo {pages} {063501}
  (\bibinfo {year} {2014}{\natexlab{c}})},\ \Eprint
  {http://arxiv.org/abs/1405.7272} {arXiv:1405.7272 [astro-ph.CO]} \BibitemShut
  {NoStop}%
%%CITATION = ARXIV:1405.7272;%%
\bibitem [{\citenamefont {Artymowski}\ and\ \citenamefont
  {Lalak}(2014)}]{Artymowski:2014gea}%
  \BibitemOpen
  \bibfield  {author} {\bibinfo {author} {\bibfnamefont {M.}~\bibnamefont
  {Artymowski}}\ and\ \bibinfo {author} {\bibfnamefont {Z.}~\bibnamefont
  {Lalak}},\ }\href {\doibase 10.1088/1475-7516/2014/09/036} {\bibfield
  {journal} {\bibinfo  {journal} {JCAP}\ }\textbf {\bibinfo {volume} {1409}},\
  \bibinfo {pages} {036} (\bibinfo {year} {2014})},\ \Eprint
  {http://arxiv.org/abs/1405.7818} {arXiv:1405.7818 [hep-th]} \BibitemShut
  {NoStop}%
%%CITATION = ARXIV:1405.7818;%%
\bibitem [{\citenamefont {Ben-Dayan}\ \emph {et~al.}(2014)\citenamefont
  {Ben-Dayan}, \citenamefont {Jing}, \citenamefont {Torabian}, \citenamefont
  {Westphal},\ and\ \citenamefont {Zarate}}]{Ben-Dayan:2014isa}%
  \BibitemOpen
  \bibfield  {author} {\bibinfo {author} {\bibfnamefont {I.}~\bibnamefont
  {Ben-Dayan}}, \bibinfo {author} {\bibfnamefont {S.}~\bibnamefont {Jing}},
  \bibinfo {author} {\bibfnamefont {M.}~\bibnamefont {Torabian}}, \bibinfo
  {author} {\bibfnamefont {A.}~\bibnamefont {Westphal}}, \ and\ \bibinfo
  {author} {\bibfnamefont {L.}~\bibnamefont {Zarate}},\ }\href {\doibase
  10.1088/1475-7516/2014/09/005} {\bibfield  {journal} {\bibinfo  {journal}
  {JCAP}\ }\textbf {\bibinfo {volume} {1409}},\ \bibinfo {pages} {005}
  (\bibinfo {year} {2014})},\ \Eprint {http://arxiv.org/abs/1404.7349}
  {arXiv:1404.7349 [hep-th]} \BibitemShut {NoStop}%
%%CITATION = ARXIV:1404.7349;%%
\bibitem [{\citenamefont {Rinaldi}\ \emph
  {et~al.}(2014{\natexlab{a}})\citenamefont {Rinaldi}, \citenamefont {Cognola},
  \citenamefont {Vanzo},\ and\ \citenamefont {Zerbini}}]{Rinaldi:2014gua}%
  \BibitemOpen
  \bibfield  {author} {\bibinfo {author} {\bibfnamefont {M.}~\bibnamefont
  {Rinaldi}}, \bibinfo {author} {\bibfnamefont {G.}~\bibnamefont {Cognola}},
  \bibinfo {author} {\bibfnamefont {L.}~\bibnamefont {Vanzo}}, \ and\ \bibinfo
  {author} {\bibfnamefont {S.}~\bibnamefont {Zerbini}},\ }\href {\doibase
  10.1088/1475-7516/2014/08/015} {\bibfield  {journal} {\bibinfo  {journal}
  {JCAP}\ }\textbf {\bibinfo {volume} {1408}},\ \bibinfo {pages} {015}
  (\bibinfo {year} {2014}{\natexlab{a}})},\ \Eprint
  {http://arxiv.org/abs/1406.1096} {arXiv:1406.1096 [gr-qc]} \BibitemShut
  {NoStop}%
%%CITATION = ARXIV:1406.1096;%%
\bibitem [{\citenamefont {Rinaldi}\ \emph
  {et~al.}(2014{\natexlab{b}})\citenamefont {Rinaldi}, \citenamefont {Cognola},
  \citenamefont {Vanzo},\ and\ \citenamefont {Zerbini}}]{Rinaldi:2014gha}%
  \BibitemOpen
  \bibfield  {author} {\bibinfo {author} {\bibfnamefont {M.}~\bibnamefont
  {Rinaldi}}, \bibinfo {author} {\bibfnamefont {G.}~\bibnamefont {Cognola}},
  \bibinfo {author} {\bibfnamefont {L.}~\bibnamefont {Vanzo}}, \ and\ \bibinfo
  {author} {\bibfnamefont {S.}~\bibnamefont {Zerbini}},\ }\href@noop {} {\
  (\bibinfo {year} {2014}{\natexlab{b}})},\ \Eprint
  {http://arxiv.org/abs/1410.0631} {arXiv:1410.0631 [gr-qc]} \BibitemShut
  {NoStop}%
%%CITATION = ARXIV:1410.0631;%%
\bibitem [{\citenamefont {Costa}\ and\ \citenamefont
  {Nastase}(2014)}]{Costa:2014lta}%
  \BibitemOpen
  \bibfield  {author} {\bibinfo {author} {\bibfnamefont {R.}~\bibnamefont
  {Costa}}\ and\ \bibinfo {author} {\bibfnamefont {H.}~\bibnamefont
  {Nastase}},\ }\href {\doibase 10.1007/JHEP06(2014)145} {\bibfield  {journal}
  {\bibinfo  {journal} {JHEP}\ }\textbf {\bibinfo {volume} {1406}},\ \bibinfo
  {pages} {145} (\bibinfo {year} {2014})},\ \Eprint
  {http://arxiv.org/abs/1403.7157} {arXiv:1403.7157 [hep-th]} \BibitemShut
  {NoStop}%
%%CITATION = ARXIV:1403.7157;%%
\bibitem [{\citenamefont {Chakravarty}\ and\ \citenamefont
  {Mohanty}(2014)}]{Chakravarty:2014yda}%
  \BibitemOpen
  \bibfield  {author} {\bibinfo {author} {\bibfnamefont {G.~K.}\ \bibnamefont
  {Chakravarty}}\ and\ \bibinfo {author} {\bibfnamefont {S.}~\bibnamefont
  {Mohanty}},\ }\href@noop {} {\  (\bibinfo {year} {2014})},\ \Eprint
  {http://arxiv.org/abs/1405.1321} {arXiv:1405.1321 [hep-ph]} \BibitemShut
  {NoStop}%
%%CITATION = ARXIV:1405.1321;%%
\bibitem [{\citenamefont {Chiba}\ and\ \citenamefont
  {Yamaguchi}(2008)}]{Chiba:2008ia}%
  \BibitemOpen
  \bibfield  {author} {\bibinfo {author} {\bibfnamefont {T.}~\bibnamefont
  {Chiba}}\ and\ \bibinfo {author} {\bibfnamefont {M.}~\bibnamefont
  {Yamaguchi}},\ }\href {\doibase 10.1088/1475-7516/2008/10/021} {\bibfield
  {journal} {\bibinfo  {journal} {JCAP}\ }\textbf {\bibinfo {volume} {0810}},\
  \bibinfo {pages} {021} (\bibinfo {year} {2008})},\ \Eprint
  {http://arxiv.org/abs/0807.4965} {arXiv:0807.4965 [astro-ph]} \BibitemShut
  {NoStop}%
%%CITATION = ARXIV:0807.4965;%%
\bibitem [{\citenamefont {Gong}\ \emph {et~al.}(2011)\citenamefont {Gong},
  \citenamefont {Hwang}, \citenamefont {Park}, \citenamefont {Sasaki},\ and\
  \citenamefont {Song}}]{Gong:2011qe}%
  \BibitemOpen
  \bibfield  {author} {\bibinfo {author} {\bibfnamefont {J.-O.}\ \bibnamefont
  {Gong}}, \bibinfo {author} {\bibfnamefont {J.-c.}\ \bibnamefont {Hwang}},
  \bibinfo {author} {\bibfnamefont {W.-I.}\ \bibnamefont {Park}}, \bibinfo
  {author} {\bibfnamefont {M.}~\bibnamefont {Sasaki}}, \ and\ \bibinfo {author}
  {\bibfnamefont {Y.-S.}\ \bibnamefont {Song}},\ }\href {\doibase
  10.1088/1475-7516/2011/09/023} {\bibfield  {journal} {\bibinfo  {journal}
  {JCAP}\ }\textbf {\bibinfo {volume} {1109}},\ \bibinfo {pages} {023}
  (\bibinfo {year} {2011})},\ \Eprint {http://arxiv.org/abs/1107.1840}
  {arXiv:1107.1840 [gr-qc]} \BibitemShut {NoStop}%
%%CITATION = ARXIV:1107.1840;%%
\bibitem [{\citenamefont {Motohashi}\ and\ \citenamefont
  {Hu}(2015)}]{Motohashi:2015hpa}%
  \BibitemOpen
  \bibfield  {author} {\bibinfo {author} {\bibfnamefont {H.}~\bibnamefont
  {Motohashi}}\ and\ \bibinfo {author} {\bibfnamefont {W.}~\bibnamefont {Hu}},\
  }\href@noop {} {\  (\bibinfo {year} {2015})},\ \Eprint
  {http://arxiv.org/abs/1503.04810} {arXiv:1503.04810 [astro-ph.CO]}
  \BibitemShut {NoStop}%
%%CITATION = ARXIV:1503.04810;%%
\end{thebibliography}%

\end{document}